\documentclass{jpsj-suppl}
\usepackage{txfonts} 

\title{Beam Dynamics Studies and Design Optimisation of New Low Energy Antiproton Facilities}

\author{Javier \textsc{Resta-Lopez}, James R. \textsc{Hunt}, Carsten P. \textsc{Welsch}}

\inst{Department of Physics, University of Liverpool, L69 3BX, United Kingdom}

\email{jrestalo@liverpool.ac.uk}

\recdate{June 12, 2016}

\abst{Antiprotons, stored and cooled at low energies in a storage ring or at rest in traps, are highly desirable for the investigation of a large number of basic questions on fundamental interactions. This includes the static structure of antiprotonic atomic systems and the time-dependent quantum dynamics of correlated systems.

The Antiproton Decelerator (AD) at CERN is currently the world’s only low energy antiproton factory dedicated to antimatter experiments. New antiproton facilities, such as the Extra Low ENergy Antiproton ring (ELENA) at CERN and the Ultra-low energy Storage Ring (USR) at FLAIR, will open unique possibilities. They will provide cooled, high quality beams of extra-low energy antiprotons at intensities exceeding those achieved presently at the AD by factors of ten to one hundred.

These facilities, operating in the energy regime between 100 keV down to 20 keV, face several design and beam dynamics challenges, for example nonlinearities, space charge and scattering effects limiting beam life time. Detailed investigations into the low energy and long term beam dynamics have been carried out to address many of those challenges towards the design optimisation. Results from these studies are presented in this contribution, showing some examples for ELENA.}

\kword{antiprotons, low energy, accelerator optics, beam dynamics}

\begin{document}
\maketitle

\section{Introduction}

Antiprotons, stored and cooled at low energies in a storage ring or at rest in traps, are highly desirable for the investigation of a large number of basic questions on fundamental interactions, on the static structure of exotic antiprotonic atomic systems or of (radioactive) nuclei as well as on the time-dependent quantum dynamics of correlated systems. Fundamental studies include for example CPT tests by high-resolution spectroscopy of the 1s-2s transition or of the ground-state hyperfine structure of antihydrogen, as well as gravity experiments with antimatter. In addition, low-energy antiprotons are the ideal and perhaps the only tool to study in detail correlated quantum dynamics of few-electron systems in the femto and sub-femtosecond time regime.  

New facilities, currently being developed, will allow to access to extra low energies and investigate in detail many of the above questions. For example, ELENA, currently under construction at CERN, shall further decelerate the antiprotons injected from the AD at 5.3 MeV to 100 keV, with a beam population of $\sim 10^7$ cooled antiprotons. This will enable all experiments working at the AD to get lower energy, higher quality and more abundant antiproton beams, enabling the production of larger quantities of antihydrogen. A complete description of ELENA can be found in \cite{Carli1, Carli2}. Table~\ref{t1} shows some relevant beam parameters.

\begin{table}[tbh]
\caption{Basic ELENA $\bar{\textrm{p}}$ parameters.}
\label{t1}
\begin{tabular}{ll}
\hline
Kinetic energy range & 5.3~MeV--100~keV \\
Momentum range & 100~MeV$/c$--13.7~MeV$/c$\\
Intensity & $\sim 10^7 \bar{\textrm{p}}$ \\
Transverse acceptance & 75~mm$\cdot$mrad \\ 
\multicolumn{2}{l}{\bf Parameters at ejection:} \\
Number of bunches & 4 \\
$\Delta p/p$ (rms) &  $\sim 0.05\%$ \\
Bunch length (rms) & 0.33~m \\
$\epsilon_{x,y}$ (rms) & $\sim 1$~mm$\cdot$mrad\\
\hline
\end{tabular}
\end{table}

The design and optimisation of these new facilities operating in an unprecedented range of low energies, require careful single-particle tracking simulations to study optics performance and multiparticle tracking simulations to investigate beam quality and stability in the presence of beam cooling, space charge and scattering processes.  

It is also important to carry out advanced beam diagnostics R\&D to measure and control beam parameters operating with low energy and intensity. An overview of beam diagnostics for low energy antiprotons can be found in \cite{Carsten2}.

Here we focus on performance simulation studies for low energy antiproton rings, showing examples of results for ELENA. Concretely, we pay special attention to single-particle tracking simulations to optimise the lattice and evaluate important optics parameters (Sec.~\ref{secopt}) and multi-particle tracking simulations, including electron cooling and heating processes, to evaluate long-term beam life time and phase space stability (Sec.~\ref{longsimulations}). 

\section{Optics Optimisation}
\label{secopt}

Single-particle tracking simulations are essential to characterise the optics performance of all accelerator designs. It helps to evaluate the physical acceptance of the machine as well as its Dynamic Aperture (DA) in the presence of nonlinearities, which may contribute to emittance distortion and particle loss. 

There are several possible nonlinearity sources, e.g. nonlinear magnetic elements used to obtain stable operating regions (sextupoles, octupoles, dodecapoles, etc.), fringe field components of quadrupoles and dipoles, and higher order nonlinear errors in linear magnet elements. 

For the case of low energy electrostatic storage rings, such as the USR \cite{Carsten}, a complete study of beam dynamics including nonlinear terms (derived from a harmonic Fourier analysis of the electric field distribution in electrostatic deflectors) has been developed in \cite{pappas}. For magnetic rings, such as ELENA, chromaticity correction sextupoles are usually the dominant source of nonlinearities and it has to be studied in detail. 

\subsection{Chromaticity correction} 

\begin{figure}[tbh]
\centering
\includegraphics[width=7cm]{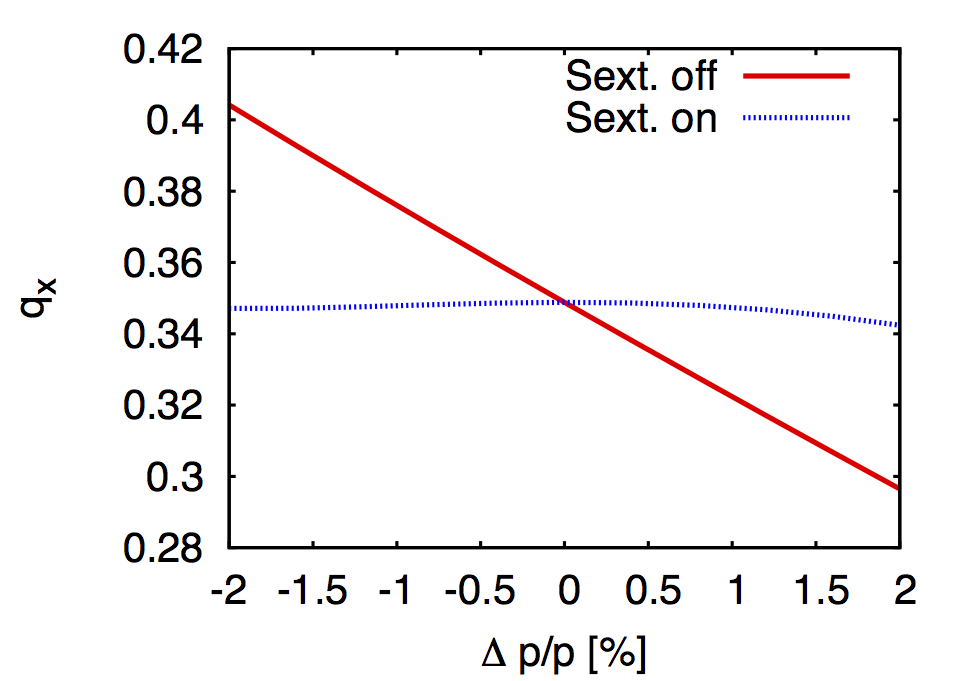}
\includegraphics[width=7cm]{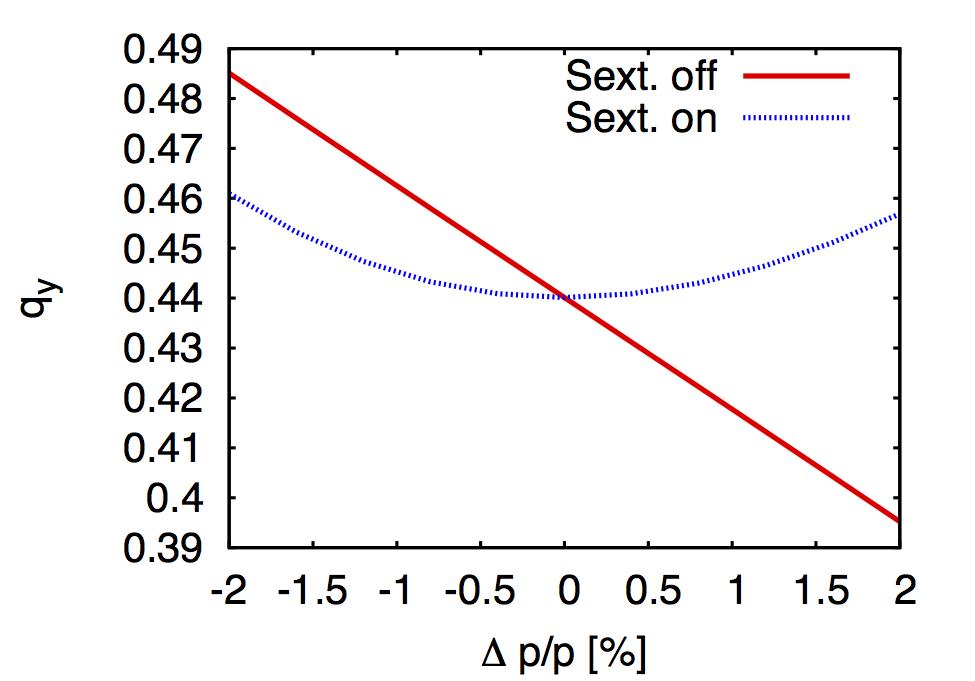}
\caption{Non-integer (fractional) part of the horizontal (left) and vertical (right) tunes versus relative momentum offset, comparing the cases of chromaticity correction sextupoles switched off and on.}
\label{f1}
\end{figure}

In ELENA two families of sextupoles, of two members each, are used to correct chromaticity. These sextupoles have been matched by means of the code MAD-X \cite{madx} to obtain zero first order chromaticity. Fig.~\ref{f1} shows the betatron tunes as a function of relative momentum offset for ELENA optics with a betatron tune working point $(Q_x, Q_y)=(2.35, 1.44)$ (the ELENA lattice presents good tunability in the range $2< Q_x <2.5$ and $1 <  Q_y < 1.5$). In this case, to correct chromaticity the required sextupole normalised magnetic strength is $k_{2}=25.8$~m$^{-3}$ and $k_{2}=-58.4$~m$^{-3}$ for each sextupole family, respectively. These strengths are relatively moderate and their impact on DA is studied in the next section.

\subsection{Dynamic Aperture}

The DA is defined as the maximum stable initial transverse amplitude in presence of nonlinearities. Following the Chirikov criterion \cite{chirikov}, the DA is determined by the onset of stochastic motion. The DA in the horizontal plane in the presence of a multipole magnet of $(2m)$th order ($m \ge 3$) can be estimated analytically \cite{gao}:

\begin{equation}
A_{x,2m}=\sqrt{2\beta_x(s)} \left( \frac{1}{m \beta^m_x(s(2m))} \right)^{1/(2(m-2))}\left( \frac{ | k_{m-1} | L}{m-1}\right)^{-1/(m-2)},
\label{eq:1}
\end{equation}

\noindent where $\beta_x(s)$ is the betatron function at the point where we want to calculate the DA, usually at the entrance of the lattice (injection), $\beta_x (s(2m))$ the betatron function at the position of the multipolar element, $L$ is the effective length of the multipolar magnet, and $k_{m-1}=(1/B_0 \rho)\partial^{m-1} B_y/\partial x^{m-1}$ is the normalised magnetic strength (defined according to the MAD-X \cite{madx} conventions). 

If the lattice presents a certain number of independent nonlinear components, i.e. in a simple case with no special phase and amplitude relations between them, the total DA can be calculated as:

\begin{equation}
A_{x,{\rm total}}=\left( \sum_{i} \frac{1}{A^2_{x, 6, i}}  + \sum_{j} \frac{1}{A^2_{x, 8, j}} + \sum_{k} \frac{1}{A^2_{x, 10, k}} + \cdots \right)^{-1/2},
\label{eq:2}
\end{equation}

\noindent where $A_{x, 6}$ stands for DA of sextupoles, $A_{x, 8}$ for octupoles, $A_{x, 10}$ for dodecapoles, and so on. 

As mentioned before, in ELENA there are just four sextupoles for chromaticity correction. For two sextupoles with $k_{2}=25.8$~m$^{-3}$ and $\beta_{x} \simeq 3.6$~m, two sextupoles with $k_{2}=-58.4$~m$^{-3}$ and $\beta_{x} \simeq 0.54$~m, and $\beta_x (s) \simeq 2$~m at the beginning and at the end of the one-turn mapping, from Eqs.~(\ref{eq:1}) and (\ref{eq:2}) we obtain $A_{x, {\rm total}} \approx 62$~mm.  For the vertical plane, in the presence of a single sextupole at position $s_i$:

\begin{equation}
A_{y, 6}=\sqrt{\frac{\beta_x(s_i)}{\beta_y(s_i)}(A^2_{x,6} - x^2)}.
\label{eq:3}
\end{equation}

Therefore, the total vertical DA can be estimated using Eq.~(\ref{eq:3}) (with $x=0$) and Eq.~(\ref{eq:2}) for $y$, i.e. $A_{y,{\rm total}}=\left( \sum_{i} 1/A^2_{y, 6, i} \right)^{-1/2}$. Knowing that $\beta_{y} \simeq 3.1$~m for the four ELENA sextupoles, then $A_{y,{\rm total}} \approx 63$~mm.  

\begin{figure}[tbh]
\centering
\includegraphics[width=7cm]{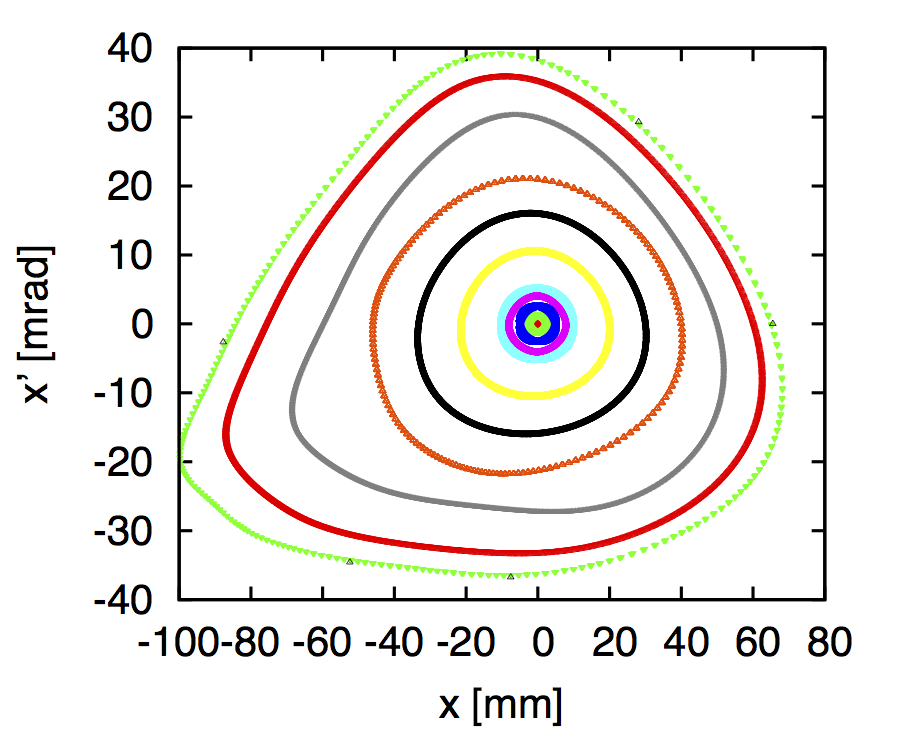}
\includegraphics[width=7cm]{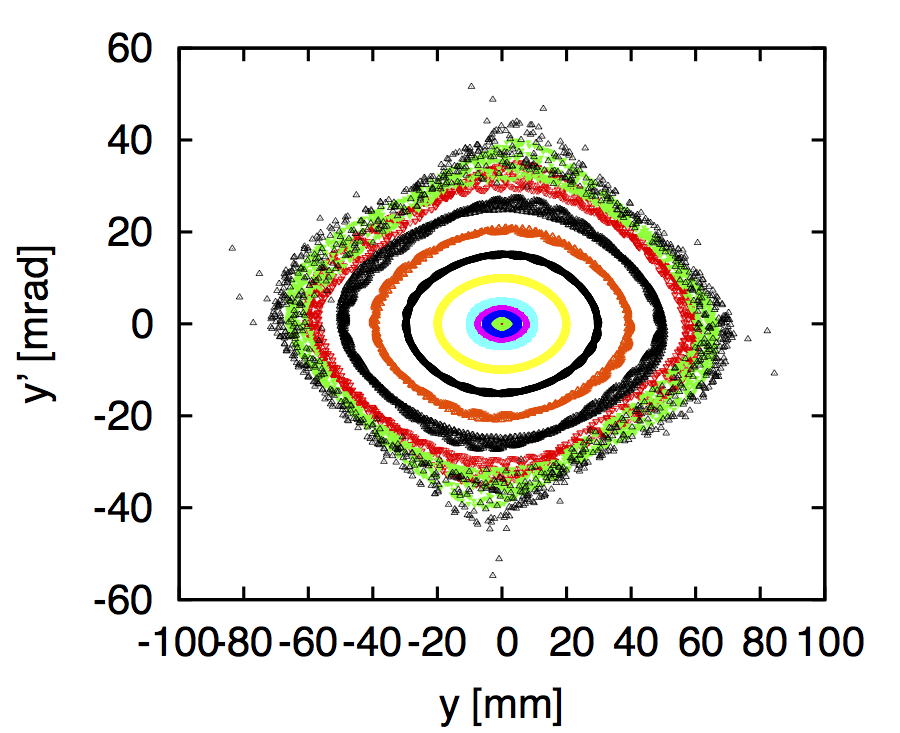}
\caption{Left: Horizontal phase space for different initial $x$ amplitudes (and zero $y$ amplitudes) tracked for 2000 turns. Right: Vertical phase space for different initial $y$ amplitudes (and zero $x$ amplitudes) tracked for 2000 turns. The chromaticity correction sextupoles are switched on. No limiting physical apertures have been considered.}
\label{f2}
\end{figure}

The above results are in reasonable agreement with multi-turn particle tracking simulations. Figure~\ref{f2} shows the transverse phase space for 2000 turns in the ELENA ring, using the PTC tracking module of the program MAD-X \cite{madx}.


Further tracking studies of ELENA (including limiting physical apertures) for different off-momentum particles have determined a momentum deviation acceptance $| \Delta p/p_0 | \lesssim 0.7\%$.

For a more complete evaluation of the DA, we need to perform numerical tracking simulations including systematic and random magnetic field imperfections. This information can be obtained from measurements for the characterisation and field mapping of each magnetic element in the ring.

\section{Long-Term Multiparticle Dynamics Simulations}
\label{longsimulations}

To evaluate the phase space evolution and life time of the beam under non-Liouvillian processes, such as electron cooling, sophisticated numerical calculations are necessary. By cooling we refer to the increase of the 6D phase space density and reduction of the 6D emittance of the beam. Taking into account other additional phenomena that may contribute to emittance growth, e.g. space charge and Intra-Beam Scattering (IBS), is also of great importance for cold and dense beams. 

To simulate the cooling process and the beam parameter evolution in presence of the above effects we have used the code BETACOOL \cite{betacool}, which allows us to perform long-term multiparticle tracking simulations, including several cooling and heating processes affecting the beam. The code BETACOOL has been benchmarked with measurements in the past, for example in the context of the low energy ion ring ELISA \cite{pappas}, giving a reasonable agreement. 

\begin{figure}[tbh]
\centering
\includegraphics[width=14cm]{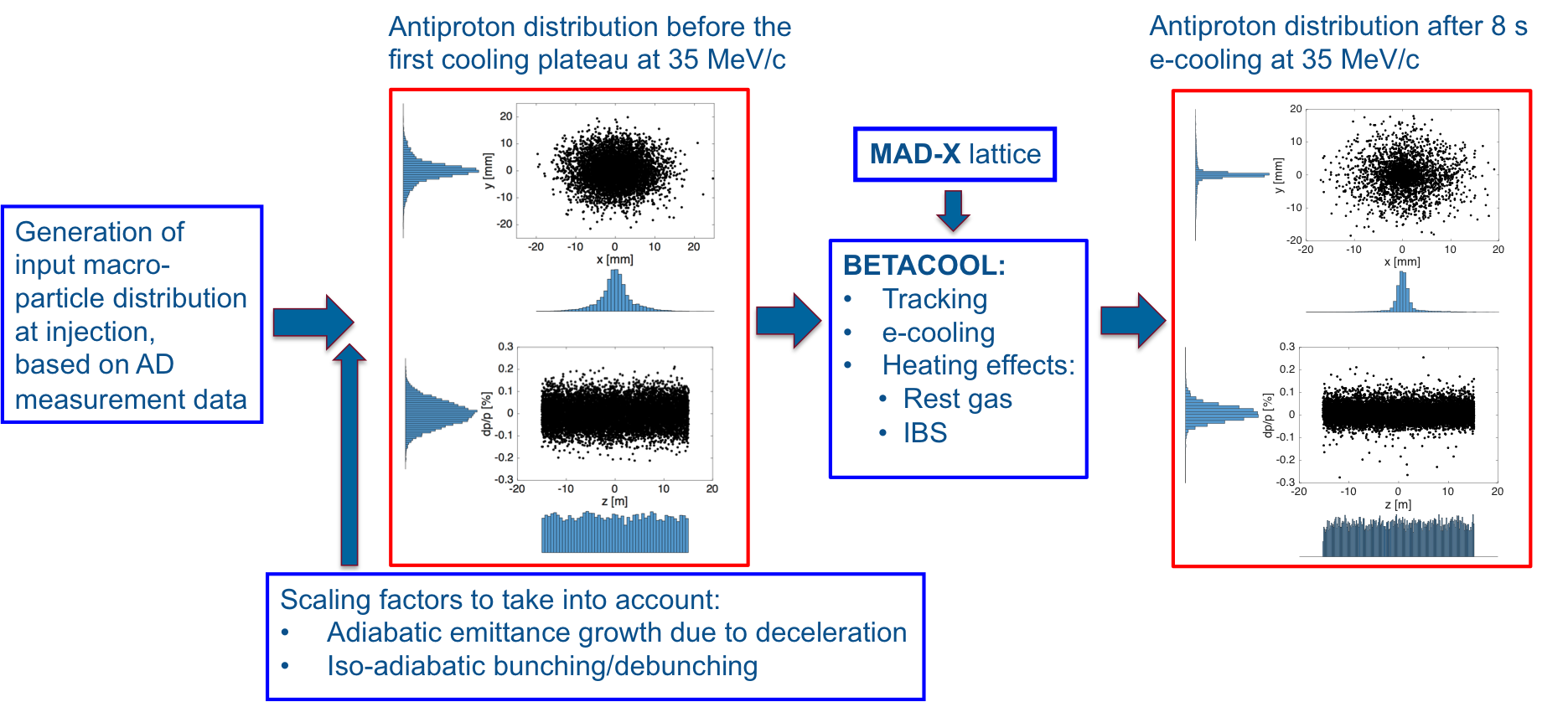}
\caption{Schematic description of the simulation sequence of the Monte Carlo multiparticle tracking in the ELENA ring applying e-cooling and different  heating scattering effects. The transverse and longitudinal beam profiles at the beginning of the first e-cooling plateau at 35 MeV$/c$ momentum is shown on the left hand side. The resulting beam profiles after 8 s e-cooling is shown on the right hand side.}
\label{f3}
\end{figure}

A schematic of our beam dynamics simulation sequence applied to the ELENA case is shown in Fig.~\ref{f3}. A particularity of these simulations is that they integrate the information of real measurements in the AD. In this way we can model more realistically antiproton beam profiles at injection from the AD to ELENA, and use this information to generate input macroparticle distributions for Monte Carlo tracking using BETACOOL. The optics lattice information is generated by the code MAD-X and read by BETACOOL. The IBS process has been introduced through the use of a self-consistent algorithm (local IBS algorithm). This model constitutes an important step towards consistent beam dynamics simulations, solving many of the issues and artefacts, such as core over-cooling of the antiproton beam, of simulations made in the past for ELENA \cite{tranquille, javier1}. 

In ELENA, e-cooling is applied at three stages of the machine cycle: after deceleration ramps, at $p=35$~MeV$/c$ and $13.7$~MeV$/c$, respectively, for a coasting beam; and during bunching prior to ejection at $13.7$~MeV$/c$. In Fig.~\ref{f3} we just show an example of beam profiles for the case of a coasting antiproton beam (simulated with $10^4$ macroparticles) before and after 8 s e-cooling. 

\begin{figure}[tbh]
\centering
\includegraphics[width=10cm]{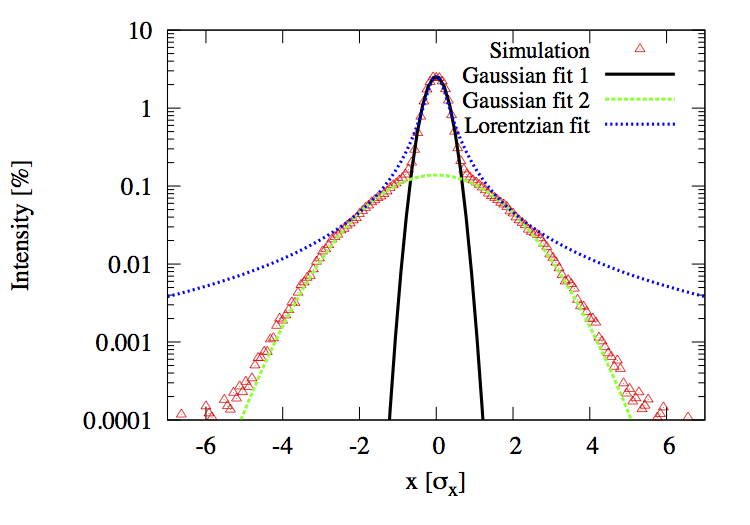}
\caption{Example of horizontal antiproton beam distribution after 8 s e-cooling, including also IBS effects, at 35 MeV/c momentum. The horizontal axis is normalised to the initial rms width. The red triangles show the result of the BETACOOL simulation. Gaussian fittings to both the dense core (black solid line) and the tails (dashed green line) are also shown. In addition, a Lorentzian function fitting has also been performed (dotted blue line).}
\label{f4}
\end{figure}

A typical core-tail beam distribution obtained after the simulation of the cooling process in presence of heating diffusion effects is shown in Fig.~\ref{f4}. It presents a dense core and long tails, which can be well represented by a bi-Gaussian function in a broad dynamic range. The central region ($-3\sigma_x < x < 3\sigma_x$) can also be well described by heavy-tailed functions, such as a Lorentz function or a L\'evy stable symmetrical distribution. A more extensive discussion can be found in \cite{javier2}.

\section{Conclusions and Outlook}

New antiproton facilities, such as ELENA at CERN and FLAIR at GSI, will open the door to a unique antimatter research programme in the coming years. For instance, in the near future the AD-ELENA complex at CERN will provide cooled, high quality beams of 100 keV kinetic energy antiprotons at intensities exceeding those achieved presently at the AD by a factor of ten to one hundred, thus improving by the same factor the efficiency of antihydrogen production.

Before commissioning and full operation of these machines at unprecedented low energy, it is important to carry out realistic beam dynamics simulations to optimise their performance and fully understand the different physics phenomena affecting beam quality and stability. 

We have started systematic studies to characterise the ELENA optics by means of analytical calculations and multi-turn particle tracking simulations. The contribution of different nonlinearities to dynamic aperture reduction is being evaluated. Here, preliminary results for ELENA sextupoles have been discussed in detail. 

Furthermore, we are currently developing a complete start-to-end simulation for the whole ELENA cycle. This includes realistic assumptions of initial particle distributions and the use of self-consistent space charge and scattering models. Of course, following the usual procedure for these kind of studies, the validation of any computation model will be determined by comparison with measurements in the commissioning stage and future operation of the machine. This will provide the necessary feedback to improve the simulation model. 

Although in this paper we have described simulations for the ELENA case, similar simulation procedures can be applied to other future low energy antiproton and ion ring designs.

\section*{Acknowledgments}

We gratefully acknowledge our colleagues from the ELENA team at CERN for very fruitful discussions.

This work is supported by the EU under Grant Agreement 624854 and the STFC Cockcroft Institute core Grant No. ST/G008248/1.


\begin{thebibliography}{9}
\bibitem{Carli1} V.~Chohan (editor) et al., ``Extra Low Energy Antiproton ring (ELENA) and its Transfer Lines, Design Report", CERN-2014-002 (2014).
\bibitem{Carli2} C.~Carli et al., ``Status of the Extra Low ENergy Ring (ELENA) Project", these Proceedings. 
\bibitem{Carsten2} C.~P.~Welsch et al., ``Beam Diagnostics for Low Energy Antiprotons", these Proceedings. 
\bibitem{Carsten} C.~P.~Welsch, ``Ultra-low energy storage ring at FLAIR", Hyperfine Interact. {\bf 213} (2012) 205.
\bibitem{pappas} A.~I.~Papash et al., ``Nonlinear and long-term beam dynamics in low energy storage rings", Phys. Rev. ST-AB \textbf{16} (2013) 060101.
\bibitem{madx} http://mad.web.cern.ch/mad/.
\bibitem{chirikov} B. V. Chirikov, ``A Universal Instability of Many-Dimensional Oscillator Systems", Phys. Rep. \textbf{52} (5) (1979) 263.
\bibitem{gao} J.~Gao, ``Analytical estimation of the dynamic apertures of circular accelerators", Nucl. Instrum. Methods Phys. Res. A \textbf{451} (2000) 545.
\bibitem{betacool} A.~Sidorin et al., ``BETACOOL program for simulations of beam dynamics in storage rings", Nucl. Instrum. Methods Phys. Res. A \textbf{558} (2006) 325.
\bibitem{tranquille} G.~Tranquile et al., ``The ELENA electron cooler: parameter choice and expected performance", Proceedings of COOL2013, WEPPO16, Murren, Switzerland, 2013.
\bibitem{javier1} J.~Resta-Lopez et al., ``Simulation studies of the beam cooling process in presence of heating effects in the Extra Low ENergy Antiproton ring (ELENA)", JINST \textbf{10} (2015) P05012.
\bibitem{javier2} J.~Resta-Lopez et al., ``Non-Gaussian beam dynamics in low energy antiproton storage rings", submitted to Nucl. Instrum. Methods Phys. Res. A. 
\end{thebibliography}
\end{document}